\documentclass[preprint]{elsarticle}

\usepackage{graphicx}

\usepackage{amssymb}
\usepackage{amsmath}

\usepackage{xspace}
\usepackage{lineno}

\begin{document}

\begin{frontmatter}



\title{Charge Symmetry Breaking in $dd\,\to\,{^4}\mathrm{He}\ \pi^0$ with WASA-at-COSY}


\author[IKPUU]{The WASA-at-COSY Collaboration\\[2ex] P.~Adlarson\fnref{fnmz}}
\author[ASWarsN]{W.~Augustyniak}
\author[IPJ]{W.~Bardan}
\author[PITue,Kepler]{M.~Bashkanov}
\author[MS]{F.S.~Bergmann}
\author[ASWarsH]{M.~Ber{\l}owski}
\author[IITB]{H.~Bhatt}
\author[Budker,Novosib]{A.~Bondar}
\author[IKPJ,JCHP]{M.~B\"uscher\fnref{fnpgi,fndus}}
\author[IKPUU]{H.~Cal\'{e}n}
% ...

\address[IKPUU]{Division of Nuclear Physics, Department of Physics and 
 Astronomy, Uppsala University, Box 516, 75120 Uppsala, Sweden}
\address[ASWarsN]{Department of Nuclear Physics, National Centre for Nuclear 
 Research, ul.\ Hoza~69, 00-681, Warsaw, Poland}
\address[IPJ]{Institute of Physics, Jagiellonian University, ul.\ Reymonta~4, 
 30-059 Krak\'{o}w, Poland}
\address[PITue]{Physikalisches Institut, Eberhard--Karls--Universit\"at 
 T\"ubingen, Auf der Morgenstelle~14, 72076 T\"ubingen, Germany}
\address[Kepler]{Kepler Center for Astro and Particle Physics, Eberhard Karls 
 University T\"ubingen, Auf der Morgenstelle~14, 72076 T\"ubingen, Germany}
\address[MS]{Institut f\"ur Kernphysik, Westf\"alische Wilhelms--Universit\"at 
 M\"unster, Wilhelm--Klemm--Str.~9, 48149 M\"unster, Germany}
\address[ASWarsH]{High Energy Physics Department, National Centre for Nuclear 
 Research, ul.\ Hoza~69, 00-681, Warsaw, Poland}
\address[IITB]{Department of Physics, Indian Institute of Technology Bombay, 
 Powai, Mumbai--400076, Maharashtra, India}
\address[Budker]{Budker Institute of Nuclear Physics of SB RAS, 11~akademika 
 Lavrentieva prospect, Novosibirsk, 630090, Russia}
\address[Novosib]{Novosibirsk State University, 2~Pirogova Str., Novosibirsk, 
 630090, Russia}
\address[IKPJ]{Institut f\"ur Kernphysik, Forschungszentrum J\"ulich, 52425 
 J\"ulich, Germany}
\address[JCHP]{J\"ulich Center for Hadron Physics, Forschungszentrum J\"ulich, 
 52425 J\"ulich, Germany}
% ...

\cortext[coau]{Corresponding author}

\fntext[fnmz]{present address: Institut f\"ur Kernphysik, Johannes 
 Gutenberg--Universit\"at Mainz, Johann--Joachim--Becher Weg~45, 55128 Mainz, 
 Germany}
\fntext[fnpgi]{present address: Peter Gr\"unberg Institut, PGI--6 Elektronische 
 Eigenschaften, Forschungszentrum J\"ulich, 52425 J\"ulich, Germany}
\fntext[fndus]{present address: Institut f\"ur Laser-- und Plasmaphysik, 
 Heinrich--Heine Universit\"at D\"usseldorf, Universit\"atsstr.~1, 40225 
 D\"usseldorf, Germany}
\fntext[fnbe]{present address: Albert Einstein Center for Fundamental Physics,
 Universit\"at Bern, Sidlerstrasse~5, 3012 Bern, Switzerland}
\fntext[fnac]{present address: III.~Physikalisches Institut~B, Physikzentrum, 
 RWTH Aachen, 52056 Aachen, Germany}
\fntext[fnde]{present address: Department of Physics and Astrophysics, 
 University of Delhi, Delhi--110007, India}
\fntext[fnsh]{present address: Department of Physics and Astronomy, University 
 of Sheffield, Hounsfield Road, Sheffield, S3 7RH, United Kingdom}

\newpage

\begin{abstract}
Charge symmetry breaking (CSB) observables are a suitable experimental tool
to examine effects induced by quark masses on the nuclear level. Previous high precision data 
from TRIUMF and IUCF are currently used to develop a consistent
description of CSB within the framework of chiral perturbation theory.
In this work the experimental studies on the reaction
$\mathrm{dd}\to\mathrm{{^4}He\pi^0}$ have been extended towards higher excess 
energies in order to provide information on the contribution of 
$p$-waves in the final state. For this, an exclusive measurement has been carried out 
at a beam momentum of $p_{d}=$~1.2~GeV/c using the WASA-at-COSY
facility. The total cross section amounts to
$\sigma_\mathrm{tot} = (118 \pm 18_\mathrm{stat}\pm 13_\mathrm{sys} \pm 8_\mathrm{ext})\,\mathrm{pb}$
and first data on the differential cross section are consistent with $s$-wave
pion production.

\end{abstract}


\end{frontmatter}


\section{Introduction}
\label{}

Within the Standard Model there are two sources of isospin
violation\footnote{Ignoring tiny effects induced by the
electro-weak sector.}, namely the electro-magnetic interaction and the differences
in the masses of the lightest quarks~\cite{Weinberg1977,Gasser1982}.
Especially in situations where one is able to disentangle these two
sources, the observation of isospin violation in hadronic reactions is a
direct window to quark mass
ratios~\cite{Gasser1982,Miller1990,Leutwyler1996}.

The effective field theory for the Standard Model in the MeV range is
chiral perturbation theory (ChPT). It maps all symmetries of the Standard Model onto hadronic
operators --- their strength then needs to be fixed either from experiment
or from lattice QCD calculations. At leading order the only parameters are
the pion mass and the pion decay constant which are the basis for a series
of famous low energy theorems in hadron--hadron scattering (see, for example,
Ref.~\cite{Weinberg1966}). Although at subleading orders the number of a priori
unknown parameters increases, the theory still provides non-trivial links
between different operators. A very interesting example is the close link 
between the quark mass induced proton-neutron mass
difference, $\Delta M_{pn}^{\rm qm}$, and, at leading order, isospin
violating $\pi N$ scattering, the Weinberg term.
In general, it is difficult to get access to quark mass effects in low energy hadron physics:
by far the largest isospin violating effect is the
pion mass difference, which also drives the spectacular energy
dependence of the $\pi^0$--photoproduction amplitude near threshold (see
Ref.~\cite{Bernard2005} and references therein).  Thus, it is important to use observables where the pion mass
difference does not contribute. An example are charge symmetry breaking (CSB)
observables --- charge symmetry is an isospin rotation by 180 degrees that exchanges
up and down quarks --- as the pion
mass term is invariant under this rotation.
For this case, the impact of soft photons has been studied
systematically~\cite{Meissner1998,Muller1999,Fettes2001,Hoferichter2009,Gasser2002}
and can be controlled. Already in 1977 Weinberg predicted a huge effect (up to 30\%
difference in the scattering lengths for $p\pi^0$ and $n\pi^0$) of
CSB in $\pi^0 N$ scattering~\cite{Weinberg1977} (see also Ref~\cite{Baru2011} for 
the recent extraction of these quantities from pionic atoms data).

While the $\pi^0p$
scattering length might be measurable in polarized neutral pion photoproduction very near
threshold~\cite{Bernstein1998}, it is not possible to measure the $n\pi^0$ channel.
As an alternative access to CSB pion--nucleon scattering
in Ref.~\cite{VanKolck2000}
it was suggested to use $NN$ induced pion production instead. There have been two successful
measurements of corresponding CSB observables, namely a measurement of $A_{fb}(pn\to               
d\pi^0)$~\cite{Opper2003} --- the forward-backward asymmetry in $pn\to{}d\pi^0$ ---
as well as of the total cross section of $dd\to {^4}\mathrm{He} \pi^0$
close to the reaction threshold~\cite{Stephenson2003}.

The first experiment was analyzed using  ChPT in Ref.~\cite{Filin2009} (see also Ref.~\cite{Bolton2010}), where it was
demonstrated that  $A_{fb}(pn\to d\pi^0)$ is directly proportional to
 $\Delta M_{pn}^{\rm qm}$, while the effect of  $\pi -\eta$ mixing,
previously believed to completely dominate this CSB
observable~\cite{Niskanen1999}, was shown to be subleading.
 The value for  $\Delta M_{pn}^{\rm qm}$ extracted 
turned out to be consistent with other, direct calculations of
this part based on dispersive analyses~\cite{Gasser1982,Cottingham1963,
{Walker-Loud2012}} and from lattice. See Ref.~\cite{Walker-Loud2014} for the
latest review. In order to cross check the systematics and to 
eventually reduce the uncertainties, additional experimental information
needs to be analyzed.

\begin{figure*}
\includegraphics[width=0.33\textwidth]{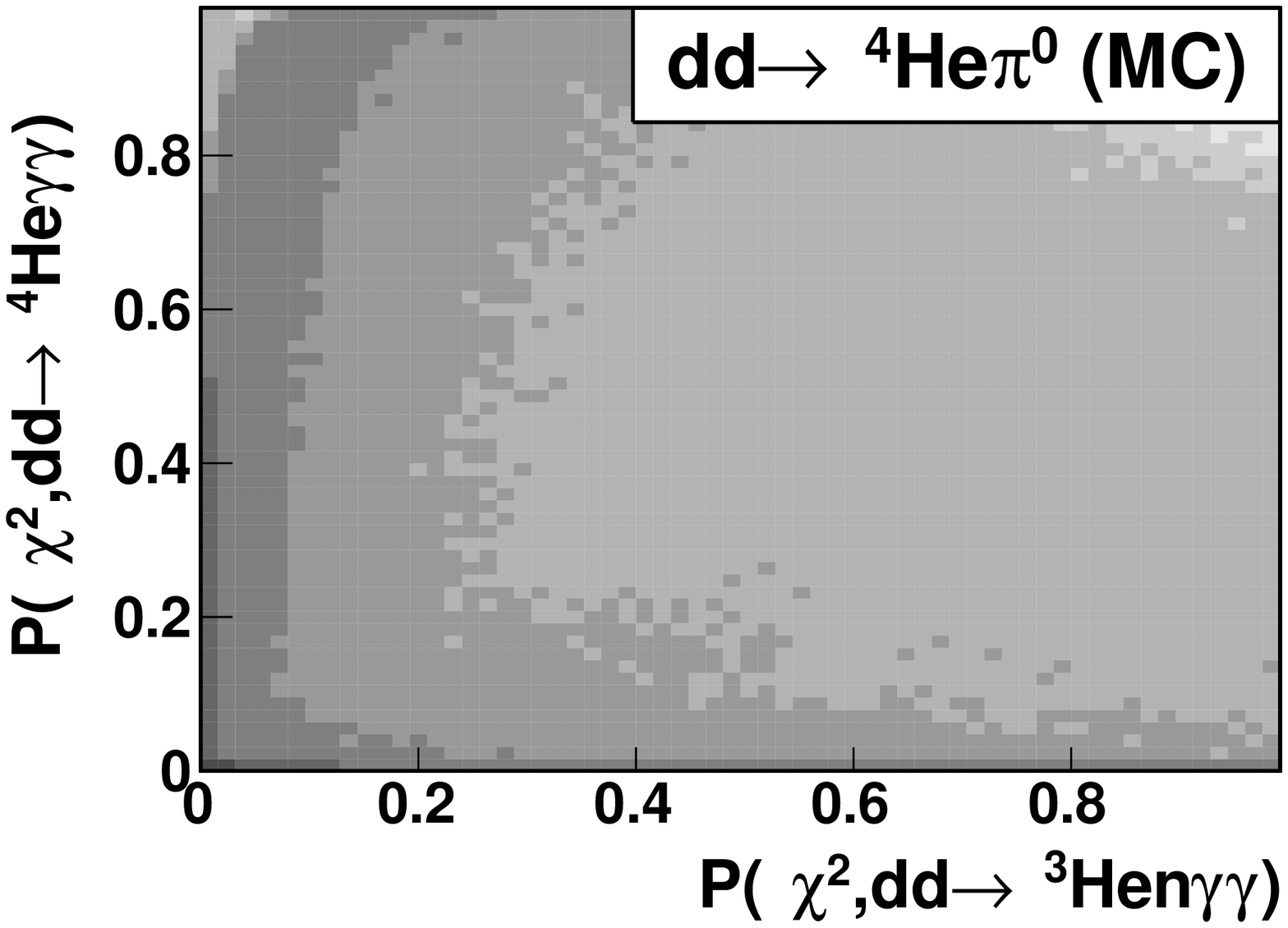}\hfill
\includegraphics[width=0.33\textwidth]{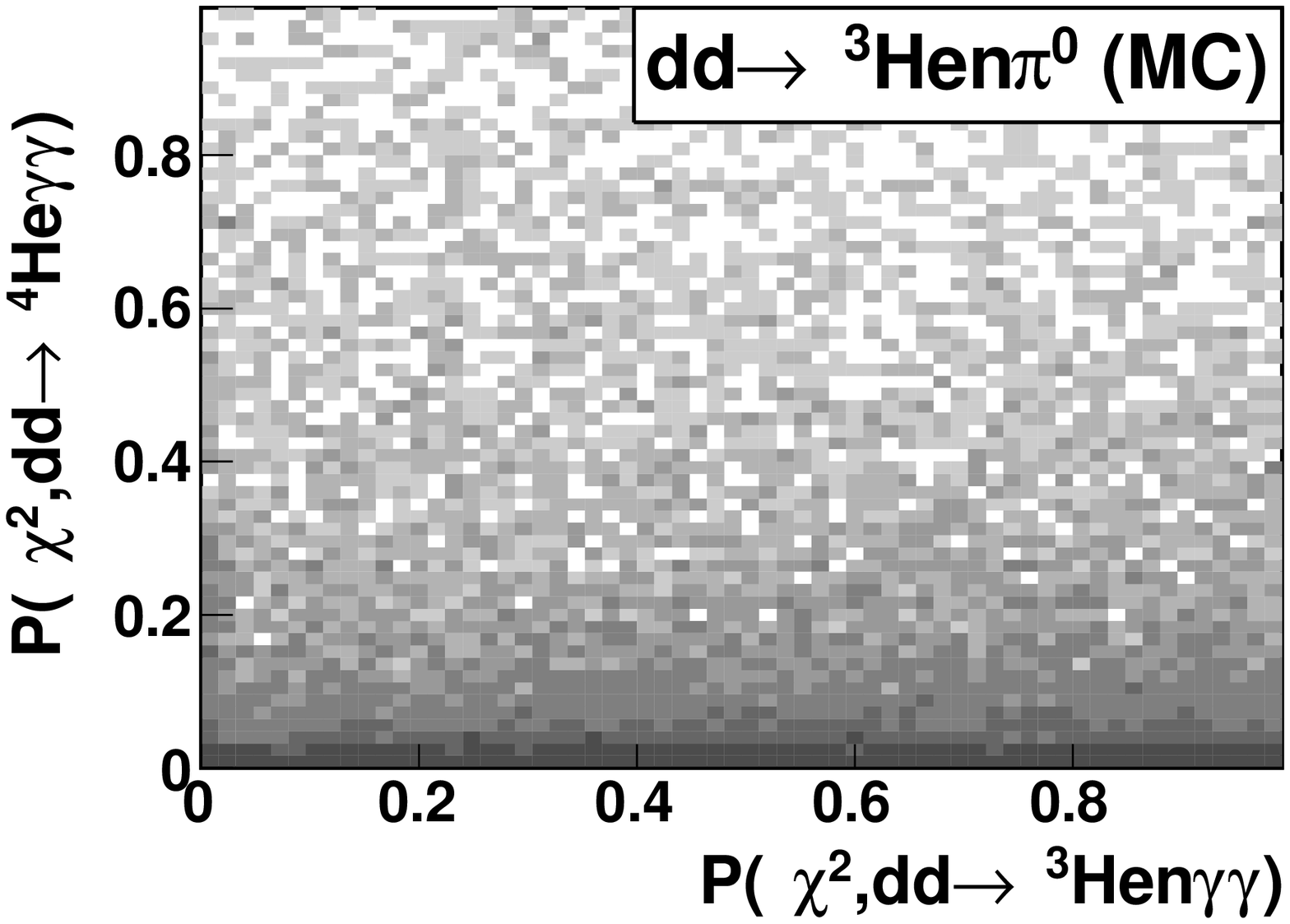}\hfill
\includegraphics[width=0.33\textwidth]{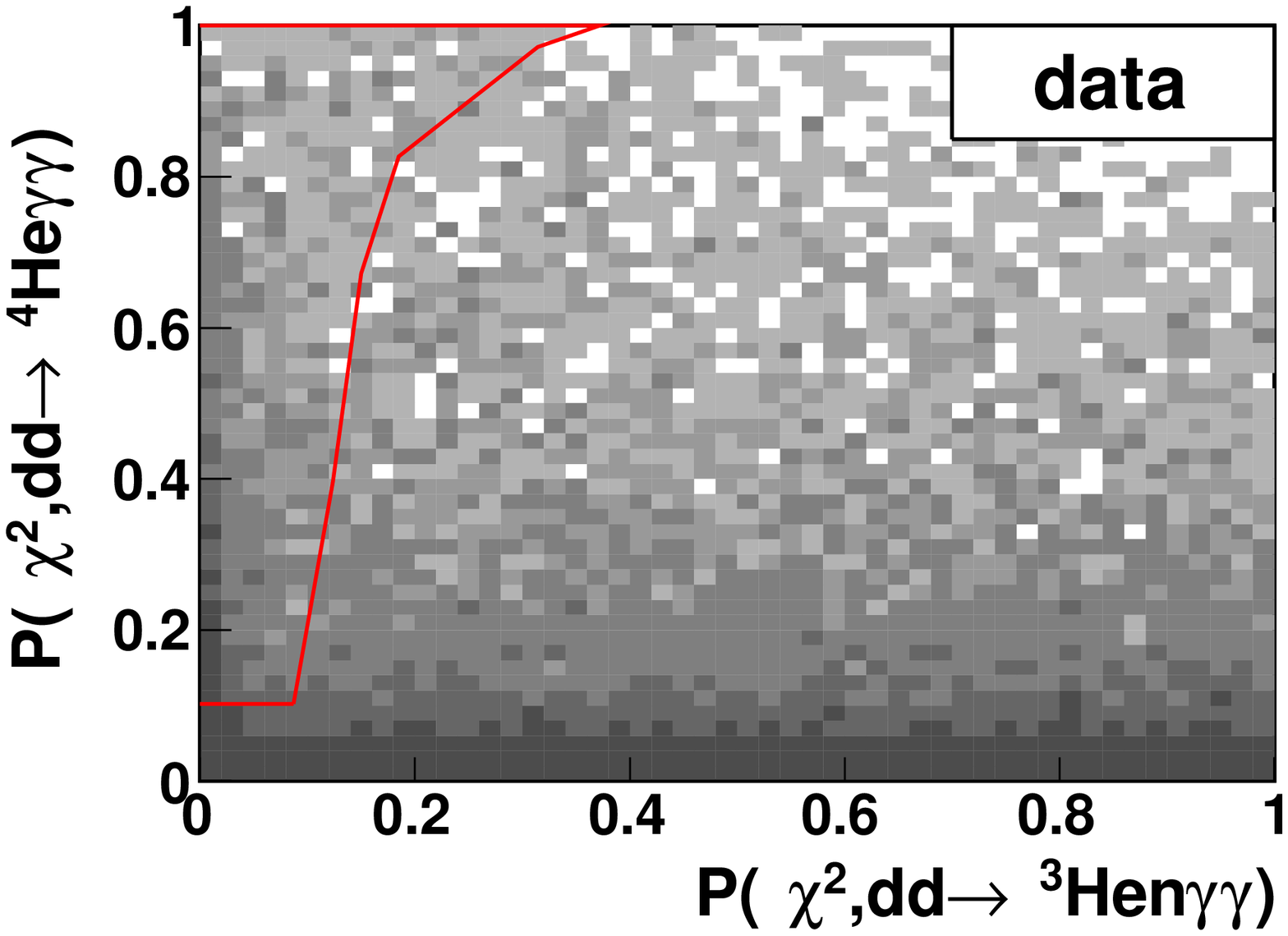}
\caption{
\label{prob}
Cumulative probability distributions from the kinematic fit used for event selection plotted 
as probability for the $\mathrm{{^4}He}$ hypothesis versus the probability for the 
$\mathrm{{^3}He}$ hypothesis. Left: distribution for Monte-Carlo simulated
signal events for $dd\to {^4}\mathrm{He}\pi^0$, middle: distribution for Monte-Carlo simulated 
events for $dd\to {^3}\mathrm{He} n \pi^0$, right: distribution for data and the applied probability cut.}
\end{figure*}

The first theoretical results for $dd\to {^4}\mathrm{He}\pi^0$ are presented
in~\cite{Gardestig2004,Nogga2006}.  The studies show that the relative
importance of the various charge symmetry breaking mechanisms is very different
compared to $pn\to{}d\pi^0$. For example, soft photon exchange may significantly
enhance the cross sections for $dd\to{^4}\mathrm{He}\pi^0$~\cite{Lahde2007}.
Furthermore, a significant sensitivity of the results to the
nuclear potential model was reported in Ref.~\cite{Fonseca2009}, which called
for a simultaneous analysis
of CSB in the $NN$ scattering length and in $dd\to{^4}\mathrm{He}\pi^0$~\cite{Fonseca2009}. Thus, 
as part of a consistent investigation of CSB in the two nucleon sector,
$pn\to{}d\pi^0$ and $dd\to{^4}\mathrm{He}\pi^0$ 
should help to further constrain the relevant CSB mechanisms.

The main challenge in the calculation of  $dd\to {^4}\mathrm{He}\pi^0$ is to get
theoretical control over the initial state interactions: high
accuracy wave functions are needed for $dd\to 4N$ in low partial waves
at relatively high energies. One prerequisite to control this is
the earlier WASA-at-COSY measurement of $dd\to {^3}\mathrm{He} n \pi^0$~\cite{WASAdd2Henpi}, 
which is allowed by charge symmetry and partially shares the same initial state
as $dd\to {^4}\mathrm{He} \pi^0$.
In addition, higher partial waves are predicted to be very sensitive
to the CSB $NN\to N\Delta$ transition potential that is difficult to access in other
reactions. In leading order in chiral perturbation theory this potential 
is known. Thus, a measurement of, for example, $p$-waves provides an additional,
non-trivial test of our current understanding of isospin violation in
hadronic systems. Future theoretical CSB studies for 
$dd\to {^4}\mathrm{He} \pi^0$ can be
based on recent developments in effective field
theories for few--nucleon systems~\cite{Epelbaum2008ga} as well as
for the reaction $NN\to NN\pi$~\cite{Filin2012,Filin2013,review},
thus promising  a model-independent analysis of the data.

While the previous measurements of $dd\to {^4}\mathrm{He}\pi^0$ close
to reaction threshold were limited to the total cross section~\cite{Stephenson2003},
in order to extract constraints on higher partial waves any new measurement at higher excess 
energies in addition has to provide information on the differential cross section.
For this, an exclusive measurement detecting the ${^4}\mathrm{He}$ ejectile as well
as the two decay photons of the $\pi^0$ has been carried out utilizing the same setup
used for $dd\to {^3}\mathrm{He} n \pi^0$~\cite{WASAdd2Henpi}. The latter reaction was also used
for normalization.

\section{Experiment}
The experiment was carried out at the Institute for Nuclear Physics of the Forschungszentrum
J\"ulich in Germany using the Cooler Synchrotron COSY \cite{Maier97} together with the
WASA detection system \cite{wasa}. For the measurement of $dd \to \mathrm{{^4}He}\pi^{0}$ at 
an excess energy of $Q\approx\,$60~MeV a deuteron beam with a momentum of 1.2~GeV/c was 
scattered on frozen deuterium pellets provided by an internal pellet target. The $\mathrm{{^4}He}$ 
ejectile and the two photons from the $\pi^{0}$ decay were detected by the Forward Detector and 
the  Central Detector of the WASA facility, respectively. The experimental setup and trigger
conditions were the same as described in Ref.~\cite{WASAdd2Henpi}. 

\section{Data Analysis}
The basic analysis leading to event samples with one helium and two photons in final state
follows the strategy used for $dd \to \mathrm{{^3}He}n\pi^0$ outlined in Ref.~\cite{WASAdd2Henpi}. 
Compared to this reaction, however, the charge symmetry breaking reaction $dd \to \mathrm{{^4}He}\pi^{0}$ has 
a more than four orders of magnitude smaller cross section. The only other channel with $\mathrm{{^4}He}$ 
and two photons in final state is the double radiative capture reaction $dd \to \mathrm{{^4}He}\gamma\gamma$.
The cross sections for both reactions are not enough large to provide a visual 
signature for $\mathrm{{^4}He}$ in the previously used $\Delta E-\Delta E$ plots from the Forward Detector. 
Thus, all $\mathrm{{^3}He}$ and $\mathrm{{^4}He}$ candidates 
together with the two photons have been tested against the hypotheses $dd \to \mathrm{{^4}He}\gamma\gamma$ 
(``$\mathrm{{^4}He}$ hypothesis'') and $dd \to \mathrm{{^3}He}n\gamma\gamma$ (``$\mathrm{{^3}He}$ hypothesis'')
by means of a kinematic fit. Besides the overall energy and momentum conservation no other constraints 
have been included. Especially, there is no constraint on the invariant mass of the two photons in order to leave 
a decisive missing-mass plot and not to introduce a fake $\mathrm{{^4}He}\pi^{0}$ signal. 

For final event classification the cumulative probabilities $P(\chi^2, n.d.f.)$ for the two hypotheses
have been plotted as probability for the $\mathrm{{^4}He}$ hypothesis versus the probability for the 
$\mathrm{{^3}He}$ hypothesis (see Fig.~\ref{prob}). The data (right plot) have been compared 
to Monte-Carlo generated samples of $dd \to \mathrm{{^4}He}\pi^{0}$ events (left plot) and 
$dd \to \mathrm{{^3}He}n\pi^{0}$ events (middle plot). 
Events originating from $dd \to \mathrm{{^4}He}\pi^{0}$ populate the low 
probability region for the $\mathrm{{^3}He}$ hypothesis and form a uniform distribution for the 
$\mathrm{{^4}He}$ hypothesis. As there is no pion constraint in the fit, 
events from the double radiative capture reaction show the same signature. 
For $dd \to \mathrm{{^3}He}n\pi^{0}$  the situation is opposite. 
The indicated cut is based on the Monte-Carlo simulations, but has been
optimized by maximizing the statistical significance of the $\pi^0$ signal in final missing mass plot.
In addition, it has been checked that the result is stable within the statistical errors against variations
of the probability cut. For the simulations the standard Geant3 \cite{Geant3} based WASA Monte-Carlo package 
has been used, which includes the full detector setup and which has already been benchmarked against a 
wide range of reactions from the WASA-at-COSY physics program. After this analysis step the contribution 
from misidentified $\mathrm{{^3}He}$ was reduced by about four orders of magnitude.
  
\begin{figure}
\begin{center}
\includegraphics[width=\columnwidth]{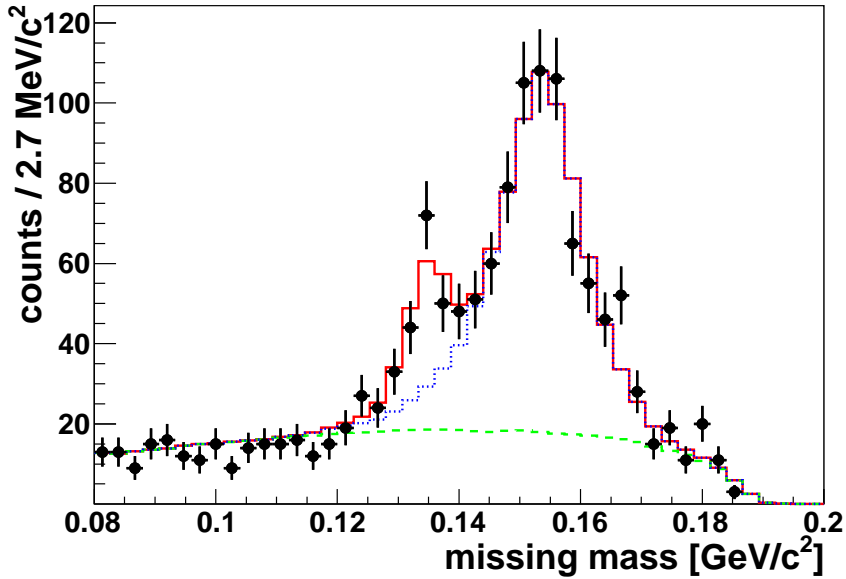}
\end{center}
\caption{
\label{total}
Missing mass plot for the reaction $dd \to \mathrm{{^4}He}X$. The different contributions fitted to the spectrum 
are double radiative capture $dd \to \mathrm{{^4}He}\gamma\gamma$ (green dashed), the reaction 
$dd \to \mathrm{{^3}He}n\pi^{0}$ (blue dotted, added) and the sum of all contributions including 
the signal (red solid).}
\end{figure}

\begin{figure*}
\begin{center}
\includegraphics[width=0.5\textwidth]{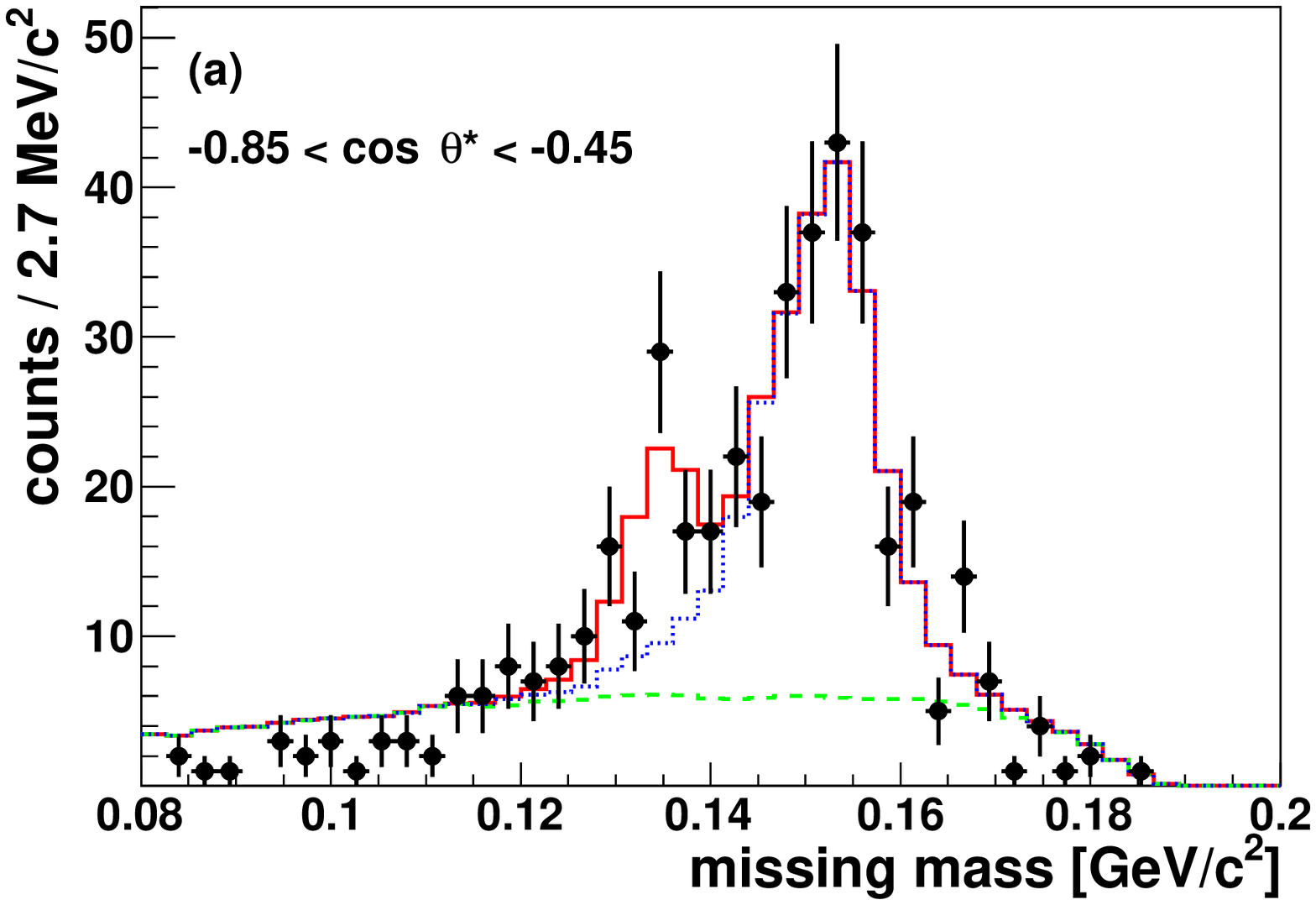}\hfill
\includegraphics[width=0.5\textwidth]{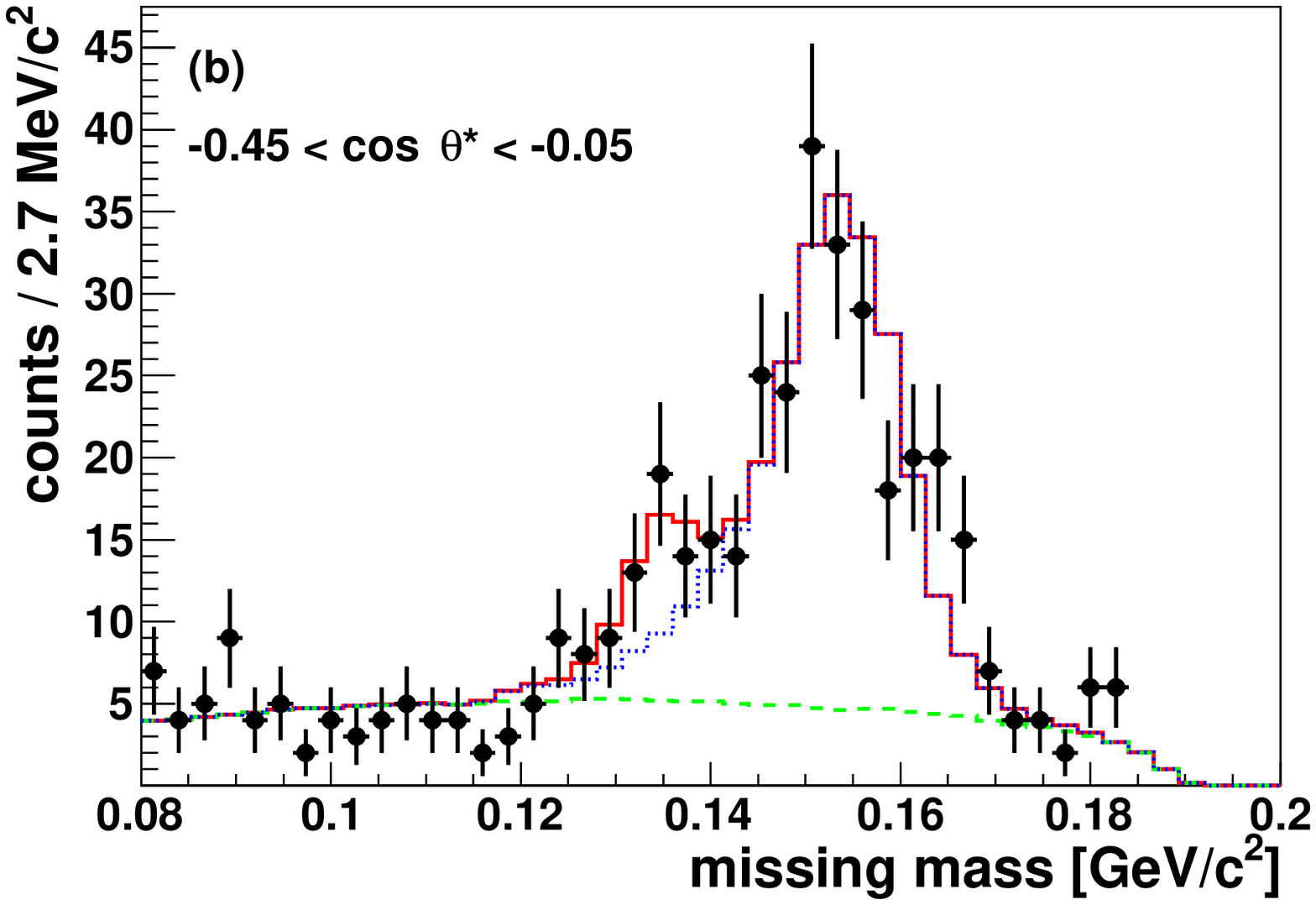}
\includegraphics[width=0.5\textwidth]{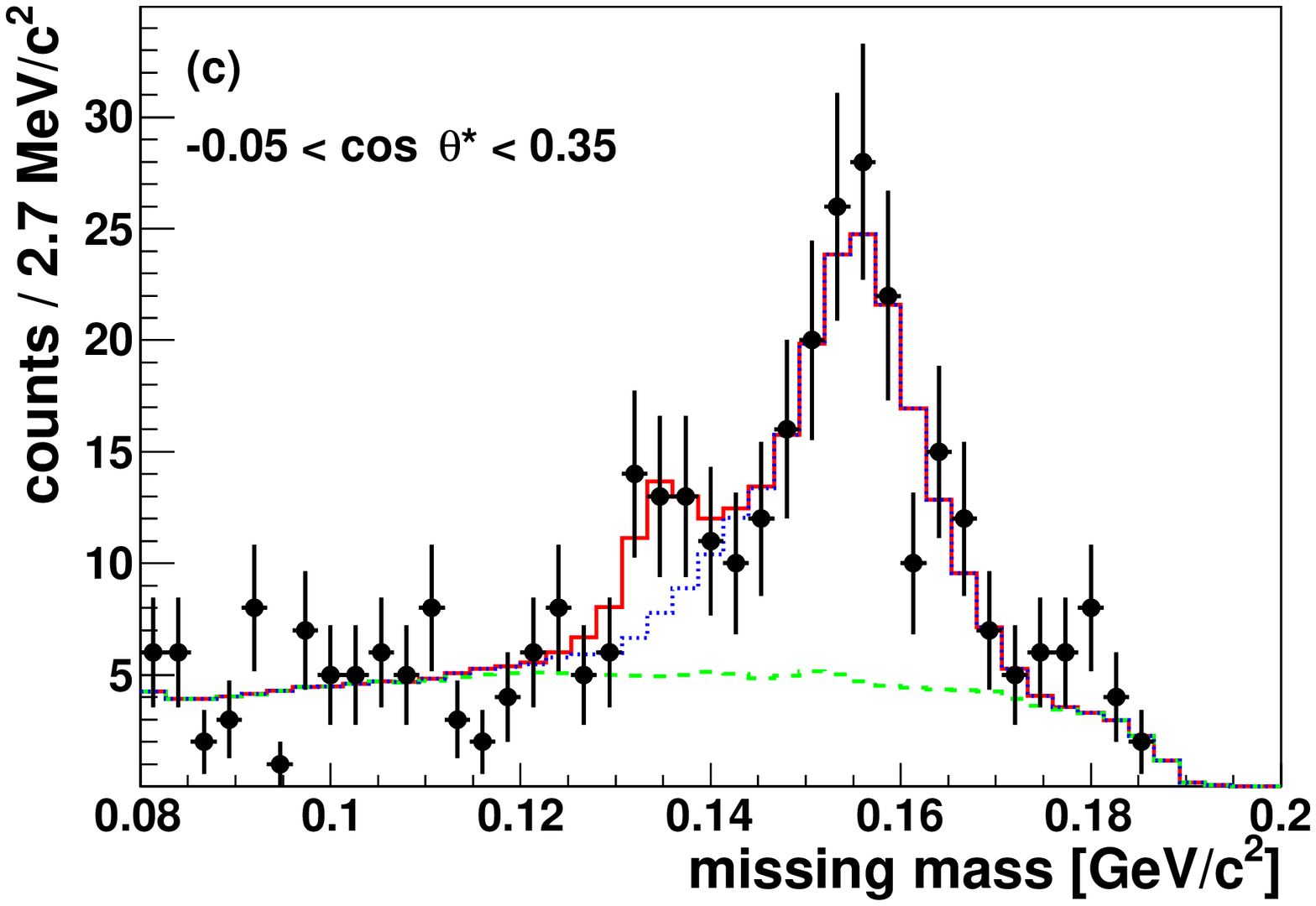}\hfill
\includegraphics[width=0.5\textwidth]{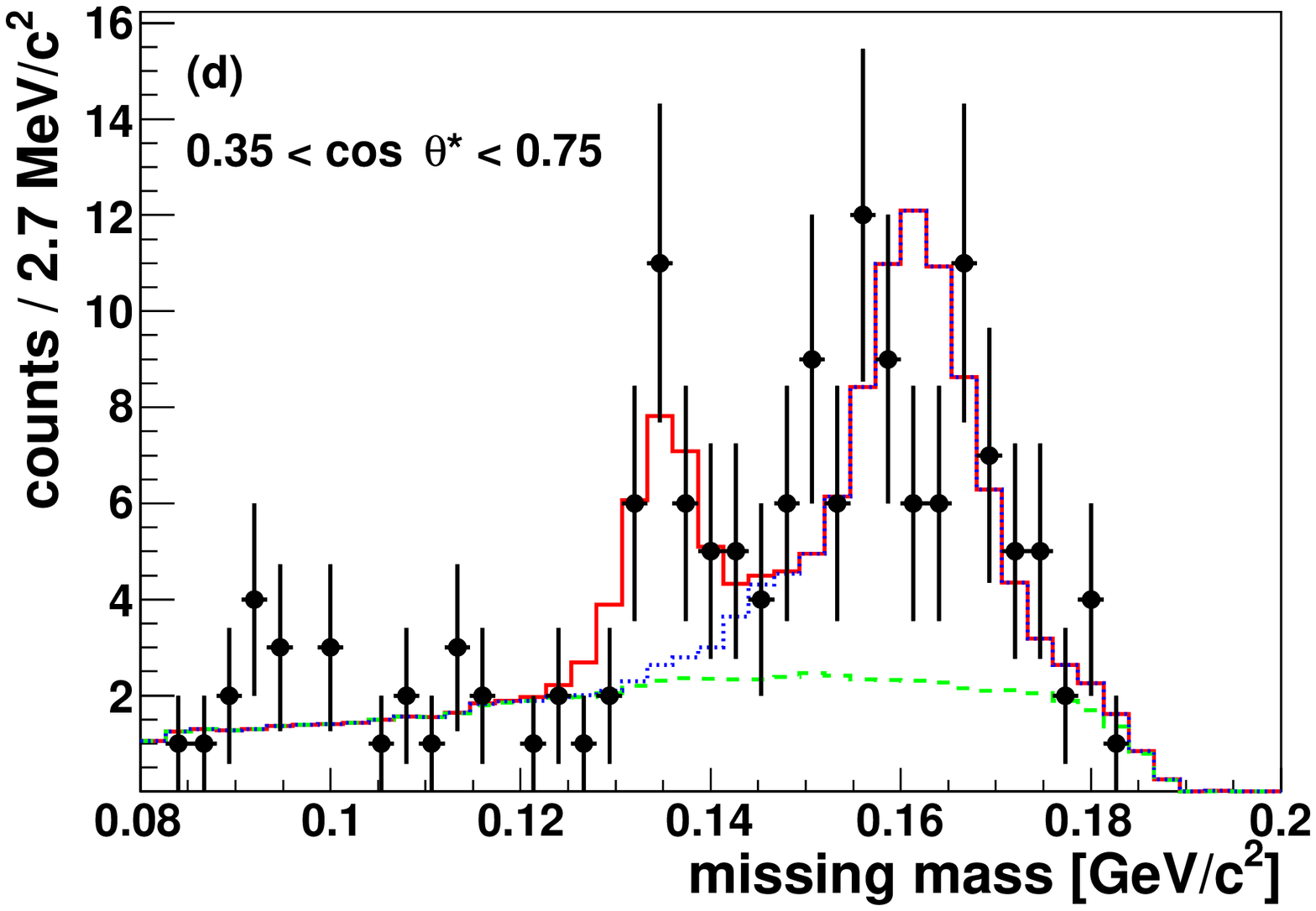}
\end{center}
\caption{\label{diff}
Missing mass plots for the four different angular bins (scattering angle of
the pion in the c.m. system). The color code for the individual contributions is the 
same as in Fig.~\ref{total}.}
\end{figure*}

In a next step, the resulting four momenta based on the fit hypothesis 
$dd \to \mathrm{{^4}He}\gamma\gamma$ 
have been used to calculate the missing mass $m_X$ in $dd \to \mathrm{{^4}He}X$ as a function of 
the center-of-mass scattering angle $\theta^*$ of the particle $X$. Figure~\ref{total} shows
a peak at the pion mass on top of a broad background. In order to extract the number of signal
events the background in the peak region has to be described and subtracted. Instead of
a (rather arbitrary) fit using a polynomial, the shape of signal and background has been   
reproduced using a composition of physics reactions with a double charged nucleus and 
two photons in the final state.  Any further 
sources of background --- physics as well as instrumental --- have already been eliminated
by the analysis steps described in Ref.~\cite{WASAdd2Henpi} and the subsequent kinematic fit.
The signal has then been extracted by fitting a linear combination of the corresponding 
Monte-Carlo generated high-statistics template distributions for the three reactions
\begin{itemize}
\item $dd \to \mathrm{{^4}He}\gamma\gamma$ (double radiative capture) using 3-body phase space (green dashed), plus
\item $dd \to \mathrm{{^3}He}n\pi^{0}$ using the model described in Ref.~\cite{WASAdd2Henpi} (blue dotted) for which
the $\mathrm{{^3}He}$ is falsely identified as $\mathrm{{^4}He}$, plus
\item $dd \to \mathrm{{^4}He}\pi^{0}$ using 2-body phase space (\emph{i.e.} plain $s$-wave, red solid).
\end{itemize}
Please note that in Fig.~\ref{total} as well as in Fig.~\ref{diff} the comulated distributions are shown, 
\emph{e.g.} the red solid curve represents the sum of all contributions.

For the differential cross section the data have been divided into four 
angular bins within the detector acceptance ($-0.85 \leq \cos\theta^* \leq 0.75$). Independent 
fits of the different contributions listed above have been performed for each bin to address possible 
anisotropies. In the course of the fit two systematic effects have been observed, which are discussed 
in the following.

First, the background originating from misidentified $\mathrm{{^3}He}$ is slightly shifted compared to
the Monte-Carlo simulations. The effect is angular dependent and is largest at forward angles. 
Possible reasons are a mismatch in the actual beam momentum, a different amount of insensitive material 
in Monte-Carlo simulations compared to the real experiment or systematic differences in the 
simulated detector response for $\mathrm{{^3}He}$ and $\mathrm{{^4}He}$ --- the limited statistics 
did not allow for a detailed study of the origin of that effect. The background stemming 
from $dd \to \mathrm{{^3}He}n\pi^{0}$ is sensitive to these effects as the energy losses from a 
(true) $\mathrm{{^3}He}$ ejectile are used for energy reconstruction of a (falsely identified) $\mathrm{{^4}He}$. 
The mismatch can be compensated by introducing an angular dependent scaling factor on the missing mass 
axis for the  $\mathrm{{^3}He}n\pi^{0}$ background, which has been included in the fit as additional
free parameter. For the angular bins from backward to forward these factors 
are $1.0$, $0.99$, $0.97$ and $0.94$, respectively. As the resulting fits describe the shape of the
data especially in the region of the pion peak, no additional systematic error has been assigned to
this effect. 

The second systematic effect concerns a mismatch in the low mass range $m\le 0.11\,\mathrm{GeV/c^2}$ 
in the most backward angular bin. According to the fit only events from the reaction $dd \to \mathrm{{^4}He}\gamma\gamma$ 
contribute in this mass region. The model used for this channel was 3-body phase space, which was not
expected to provide a perfect description. However, with the dominating background from 
$dd \to \mathrm{{^3}He}n\pi^{0}$ in a wide mass range, it is currently not possible 
to disentagle the two contributions precisely enough in order to verify any more
advanced theoretical model --- this issue will be addressed in a follow-up experiment, see below. 
Consequently, the final fit excludes the corresponding missing mass range (consistently in all angular bins). 
Based on the difference to the fit with the low mass region included a corresponding systematic uncertainty
for this effect has been assigned in the result. 

Figure~\ref{diff} shows the fitted missing mass spectra for the different bins 
in $\cos\theta^*$ together with the fit result. The chosen ansatz provides a good 
overall description of the full data set. Any tests for further systematic effects 
(according to the definition in Ref.~\cite{Barlow}), for example concerning rate effects and 
selection cuts in the basic analysis (see Ref.~\cite{WASAdd2Henpi}), did not reveal any additional 
systematic uncertainties. 

\section{Results}
\begin{figure}
\begin{center}
\includegraphics[width=\columnwidth]{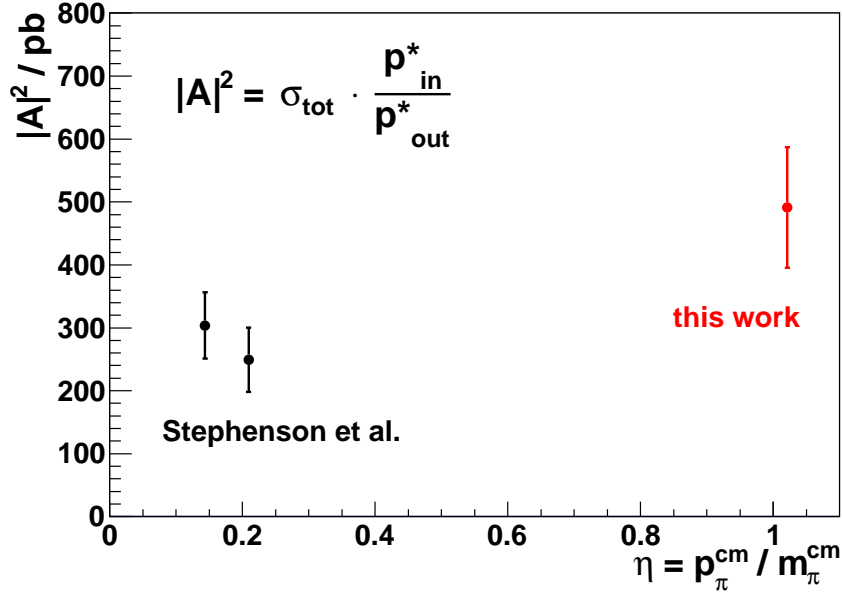}
\end{center}
\caption{
\label{amp}
Energy dependence of the reaction amplitude squared $|A|^2$. In the absence of
initial and final state interaction a constant amplitude would indicate that only
$s$-wave is contributing. The red full circle corresponds to the total cross section
given in the text.}
\end{figure}

For the acceptance correction an isotropic angular distribution has been assumed. 
For absolute normalization the reaction $dd \to \mathrm{{^3}He}n\pi^{0}$
has been used. The resulting differential cross sections 
extracted from Fig.~\ref{diff} are
\begin{align}
& \frac{\mathrm{d}\sigma}{\mathrm{d}\Omega}(-0.85{\leq}\cos\theta^*{\leq}-0.45)\!\!\!\!\!\! &{=}&\,(17.1 {\pm} 3.8 {\pm} 4.0_\mathrm{fit})\,\mathrm{pb/sr},\\
& \frac{\mathrm{d}\sigma}{\mathrm{d}\Omega}(-0.45{\leq}\cos\theta^*{\leq}-0.05)\!\!\!\!\!\! &{=}&\,(6.6  {\pm} 2.4)\,\mathrm{pb/sr},\\
& \frac{\mathrm{d}\sigma}{\mathrm{d}\Omega}(-0.05{\leq}\cos\theta^*{\leq} 0.35)\!\!\!\!\!\! &{=}&\,(5.5  {\pm} 2.2)\,\mathrm{pb/sr}, \mathrm{and}\\
& \frac{\mathrm{d}\sigma}{\mathrm{d}\Omega}(0.35{\leq}\cos\theta^*{\leq} 0.75)\!\!\!\!\!\! &{=}&\,(8.4  {\pm} 2.8)\,\mathrm{pb/sr}.
\end{align} 
In general, only statistical errors are given, except for the first bin where the uncertainty caused by 
the systematic effect in the low mass region has been included. A systematic error of $10\%$ for luminosity
determination and $7\%$ for the normalization to external data is common to all numbers.
Integrating the individual results, the (partial) total cross section within the detector acceptance amounts to
\begin{equation}
\sigma_\mathrm{tot}^\mathrm{acc} = (94 \pm 14_\mathrm{stat} \pm 10_\mathrm{sys} \pm 6_\mathrm{ext})\;\mathrm{pb}
\end{equation}
with the systematic error originating from luminosity determination and the uncertainty
from the different fit methods. The external normalization error has been propagated from
the luminosity determination for $dd \to \mathrm{{^3}He}n\pi^{0}$ (see Ref.~\cite{WASAdd2Henpi}). 
Extrapolation to full phase space by assuming an isotropic distribution yields
\begin{equation}
\sigma_\mathrm{tot} = (118 \pm 18_\mathrm{stat} \pm 13_\mathrm{sys} \pm 8_\mathrm{ext})\;\mathrm{pb}.
\end{equation}
This result can be compared with the values measured close to threshold by
dividing out phase space (see Fig.~\ref{amp}). A constant value could
be interpreted as a dominating $s$-wave, but one has to keep in mind that
the energy dependence of the formation of a $\mathrm{{^4}He}$ in the 4$N$ final state
might have some influence here, too. 

\begin{figure}
\begin{center}
\includegraphics[width=\columnwidth]{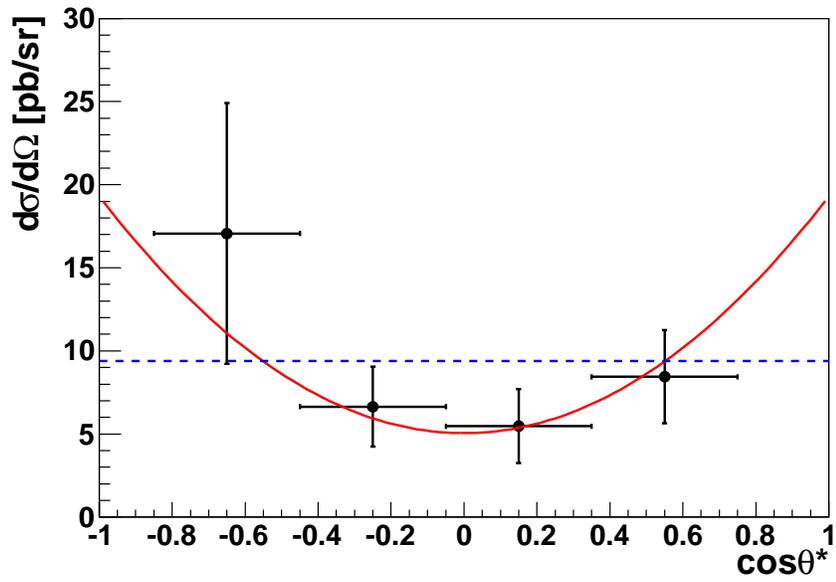}
\end{center}
\caption{
\label{diffres}
Differential cross section. The errors bars show the statistical uncertainties.
In the first bin the additional systematic uncertainty from the fit has been added 
(see text). The blue dashed line represents the total cross section given in the text assuming an isotropic distribution, 
the solid red curve shows the fit with the Legendre polynomials $P_0$ and $P_2$.}
\end{figure}

Figure~\ref{diffres} shows the differential cross section. Due to the identical particles in the initial 
state, odd and even partial waves do not interfere and the angular distribution is symmetric with 
respect to $\cos\theta^*=0$. As the $p$-wave and $s$-$d$ interference terms contribute to 
the quadratic term and the $p$-wave also adds to the constant term, the different 
partial waves cannot be directly disentangled. However, a fit including the Legendre polynomials
$P_0(\cos\theta^*)$ and $P_2(\cos\theta^*)$ --- although not excluding --- does not show any 
evidence for contributions of higher partial waves:
\begin{eqnarray}
\frac{\mathrm{d}\sigma}{\mathrm{d}\Omega} & = & (9.8 \pm 2.6)\,\mathrm{pb/sr}\,\cdot\,P_0(\cos\theta^*) \nonumber\\
                                          &   & \, +\,(9.5 \pm 7.4)\,\mathrm{pb/sr}\,\cdot\,P_2(\cos\theta^*).
\end{eqnarray}
Here, the two coefficients are strongly correlated with the correlation parameter 0.85, \emph{i.e.} the
reader should not interpret the two contributions as independent results. 

Based on the fit results a first estimate of the total cross section of $dd \to \mathrm{{^4}He}\gamma\gamma$
has been extracted assuming a homogeneous 3-body phase space. It amounts to
\begin{equation}
\sigma_\mathrm{tot}= ( 0.92 \pm 0.07_\mathrm{stat} \pm 0.10_\mathrm{\
sys} \pm 0.07_\mathrm{norm})\,\mathrm{nb}.
\end{equation}
It should be noted that this result depends on the underlying models for the reactions
$dd \to \mathrm{{^3}He}n\pi^0$ and $dd \to \mathrm{{^4}He}\gamma\gamma$. This model dependence
is not included in the given systematic error.

\section{Summary and Conclusions}

In this letter results were presented for a measurement of the charge symmetry breaking
reaction  $dd\to {^4}\mathrm{He} \pi^0$ at an excess energy of 60 MeV.  
The energy dependence of the square of the production amplitude might indicate
the on-set of higher partial waves or some unusual energy dependence of the
$s$--wave amplitude --- given the current statistical error, no  conclusion on the strength
of the higher partial waves
 is possible from the differential cross section. 

However, since within chiral perturbation theory
the leading and next-to-leading $p$-wave contribution does
not introduce any new free parameter (it is expected to be dominated
by the Delta-isobar), the data on the strength of
higher partial waves presented in this work will still provide a non-trivial 
constraint for future theoretical analyses.

The results presented here are based on a two-week run using the standard
WASA-at-COSY setup. Based on the experiences gained during this experiment
another 8 week measurement with a modified detector setup optimized for a time-of-flight
measurement of the forward going ejectiles has been performed recently. In total,
an increase of statistics by nearly a factor of 10 and significantly reduced
systematic uncertainties can be expected.  In particular, the experiment
has been designed to provide a better discrimination of background events
from $dd \to \mathrm{{^3}He}n\pi^0$.

\section{Acknowledgments}
We would like to thank the technical and administrative staff
at the For\-schungs\-zentrum J\"{u}lich, especially at the COoler SYnchrotron
COSY and at the participating institutes. This work has been supported in part by the
German Federal Ministry of
Education and Research (BMBF), the Polish Ministry of Science and Higher
Education (grants No. N N202 078135 and N N202 285938), the Polish National Science Center 
(grant No. 2011/01/B/ST2/00431), the Foundation for Polish Science (MPD),
Forschungszentrum J\"{u}lich (COSY-FFE) and the European Union
Seventh Framework Programme (FP7/2007-2013) under grant agreement No. 283286.





\section*{References}

\begin{thebibliography}{10}
\bibitem{Weinberg1977}
S.~Weinberg, Trans. N. Y. Acad. Sci. \textbf{38} (1977) 185--201.

\bibitem{Baru2011}
V.~Baru \emph{et~al.}, Phys.\ Lett.\ B {\textbf 694} (2011) 473--477; Nucl.\ Phys.\ A {\textbf 872} (2011) 69--116.

\bibitem{Gasser1982}
J.~Gasser and H.~Leutwyler, Phys. Rep. \textbf{87} (1982) 77--169.

\bibitem{Miller1990}
G.~Miller, B.~Nefkens, and I.~\v{S}laus, Phys. Rep. \textbf{194} (1990) 1--116.

\bibitem{Leutwyler1996}
H.~Leutwyler, Phys. Lett. B \textbf{378} (1996) 313--318.

\bibitem{Weinberg1966}
S.~Weinberg, Phys. Rev. Lett. \textbf{17} (1966) 616--621.

\bibitem{Bernard2005}
V.~Bernard, B.~Kubis, and U.~G. Mei{\ss}ner, Eur. Phys. J. A \textbf{25} (2005)
  419--425.

\bibitem{Meissner1998}
U.-G. Mei{\ss}ner and S.~Steininger, Phys. Lett. B \textbf{419} (1998)
  403--411.

\bibitem{Muller1999}
G.~M\"{u}ller and U.-G. Mei{\ss}ner, Nucl. Phys. B \textbf{556} (1999)
  265--291.

\bibitem{Fettes2001}
N.~Fettes and U.-G. Mei{\ss}ner, Phys. Rev. C \textbf{63} (2001) 045201.

\bibitem{Hoferichter2009}
M.~Hoferichter, B.~Kubis and U.~-G.~Mei\ss ner,
  Phys.\ Lett.\ B {\bf 678} (2009) 65.
  

\bibitem{Gasser2002}
J.~Gasser \emph{et~al.}, Eur. Phys. J. C \textbf{26} (2002) 13--34.

\bibitem{VanKolck2000}
U.~van Kolck, J.~Niskanen, and G.~Miller, Phys. Lett. B \textbf{493} (2000)
  65--72.


\bibitem{Bernstein1998}
A.~Bernstein, Phys. Lett. B \textbf{442} (1998) 20--27.

\bibitem{Opper2003}
A.~Opper \emph{et~al.}, Phys. Rev. Lett. \textbf{91} (2003) 212302.

\bibitem{Stephenson2003}
E.~Stephenson \emph{et~al.}, Phys. Rev. Lett. \textbf{91} (2003) 142302.


\bibitem{Filin2009}
A.~Filin \emph{et~al.}, Phys. Lett. B \textbf{681} (2009) 423--427.

\bibitem{Bolton2010}
D.~R. Bolton and G.~A. Miller, Phys. Rev. C \textbf{81} (2010) 014001.

\bibitem{Niskanen1999}
J.~A. Niskanen, Few-Body Syst. \textbf{26} (1999) 241--249.

\bibitem{Cottingham1963}
W.~Cottingham, Ann. Phys. (N. Y). \textbf{25} (1963) 424--432.

\bibitem{Walker-Loud2012}
A.~Walker-Loud, C.~E. Carlson, and G.~A. Miller, Phys. Rev. Lett. \textbf{108}
  (2012) 232301.

\bibitem{Walker-Loud2014}
A.~Walker-Loud,   
PoS LATTICE {\textbf 2013} (2014) 013.

\bibitem{Gardestig2004}
A.~G{\aa}rdestig \emph{et~al.}, Phys. Rev. C \textbf{69} (2004) 044606.

\bibitem{Nogga2006}
A.~Nogga \emph{et~al.}, Phys. Lett. B \textbf{639} (2006) 465--470.

\bibitem{Lahde2007}
T.~L\"{a}hde and G.~Miller, Phys. Rev. C \textbf{75} (2007) 055204.

\bibitem{Fonseca2009}
A.~C. Fonseca, R.~Machleidt, and G.~A. Miller, Phys. Rev. C \textbf{80} (2009)
  027001.

\bibitem{WASAdd2Henpi}
  P.~Adlarson {\it et al.}  (WASA-at-COSY Collaboration),
  Phys.\ Rev.\ C {\bf 88} (2013) 014004.


\bibitem{Epelbaum2008ga}
  E.~Epelbaum, H.-W.~Hammer and U.-G.~Mei{\ss}ner,
  Rev.\ Mod.\ Phys.\  {\bf 81} (2009) 1773.

\bibitem{Filin2012}
  A.~A.~Filin et al.,
  Phys.\ Rev.\ C {\bf 85} (2012) 054001.

\bibitem{Filin2013}
  A.~A.~Filin et al.,
  Phys.\ Rev.\ C {\bf 88} (2013) 064003.

\bibitem{review}
V. Baru, C. Hanhart, and F. Myhrer, Int. J. Mod. Phys. {\bf E 23} (2014) 1430004.

\bibitem{Maier97}
R.~Maier {\it et al.}, Nucl.\ Phys.\ \textbf{A626} (1997) 395c.

\bibitem{wasa}
H.~H.~Adam {\it et al.} (WASA-at-COSY Collaboration), arXiv:nucl-ex/0411038 (2004).

\bibitem{Geant3}
{\it GEANT - Detector Description and Simulation Tool}, 
CERN Program Library Long Writeup W5013.

\bibitem{Barlow}
  R.~Barlow,
  arXiv:hep-ex/0207026 (2002).


\end{thebibliography}
\end{document}